\documentclass[conference]{IEEEtran}
\IEEEoverridecommandlockouts
\usepackage{cite}
\usepackage{amsmath,amssymb,amsfonts}
\usepackage{algorithmic}
\usepackage{graphicx}
\usepackage{textcomp}
\usepackage{xcolor}
\usepackage{mathtools}
\usepackage{multirow}
\usepackage{siunitx}
\DeclarePairedDelimiter\floor{\lfloor}{\rfloor}
\def\BibTeX{{\rm B\kern-.05em{\sc i\kern-.025em b}\kern-.08em
    T\kern-.1667em\lower.7ex\hbox{E}\kern-.125emX}}
\begin{document}

\title{A Comparative Study of Analog/Digital Self-Interference Cancellation\\ for Full Duplex Radios
{\footnotesize \textsuperscript{}}
\thanks{This work was in part supported by Agency for Defense Development.}
}

\author{\IEEEauthorblockN{Jong~Woo~Kwak\IEEEauthorrefmark{1}, Min~Soo~Sim\IEEEauthorrefmark{1}, In-Woong~Kang\IEEEauthorrefmark{2}, Jong Sung~Park\IEEEauthorrefmark{2}, Jaedon~Park\IEEEauthorrefmark{2},  and
	Chan-Byoung~Chae\IEEEauthorrefmark{1}}
\IEEEauthorblockA{\IEEEauthorrefmark{1}School of Integrated Technology, Yonsei Institute of Convergence Technology, Yonsei University, Korea\\
	Email: \{kjw8216, simms, cbchae\}@yonsei.ac.kr\\
	\IEEEauthorrefmark{2}Agency for Defense Development, Korea\\
	Email: \{iwkang, jspark61, jaedon2\}@add.re.kr}}
\maketitle

\begin{abstract}
Self-interference (SI) is the main obstacle to full-duplex radios. To overcome the SI, researchers have proposed several analog and digital domain self-interference cancellation (SIC) techniques. How well the digital cancellation works depends on the results of analog cancellation. Therefore, to analyze overall SIC performance, one should do so in an integrated manner. 
In this paper, we build a simulator that can analyze the performance of analog and digital SIC techniques. Through this simulator, we can analyze the overall SIC performance within various system parameters such as the resolution of an analog-to-digital converter (ADC) and/or nonlinearity of a power amplifier (PA). With our simulator, we expect that configurations and tuning algorithms of an active analog canceller can be optimized before real hardware implementation. \\
\end{abstract}


\section{Introduction}

As demand continues to grow for increasing spectral efficiency and data rates, researchers have discovered a highly promising technology for 5G wireless communications---full-duplex~\cite{5G}.
The main challenge in a full-duplex radio is self-interference (SI)---a phenomenon where a transmit signal is received by its own receiver.
SI significantly degrades the signal-of-interest unless it can be canceled through self-interference cancellation (SIC), and several SIC methods have been proposed to solve this problem. 
 
In typical full-duplex systems~\cite{prototype,Sachin2013Full,Survey}, SIC is done in both analog and digital domains. Analog cancellation first suppresses the SI. 
Analog cancellation can be categorized into two classes---passive and active. In the former, an RF component suppresses the SI at the propagation. In the latter, the SI is regenerated by adaptive RF components and destructively added to the received signal.
An adaptive circuit is generally used for the active analog cancellation~\cite{practical,Sachin2013Full,lincoln}. Its cost includes additional power and computational overhead.
A crucial issue then is designing an efficient circuit structure and tuning algorithm. 

Remaining SI from analog cancellation is next mitigated by digital cancellation. Digital cancellation is based on a channel estimation of the residual SI channel after analog cancellation. Therefore, digital cancellation is dependent on the results of analog cancellation. In previous studies on digital cancellation, researchers have signified the analog cancellation in the simulation just as linear attenuation~\cite{asilomar,nonlinear3}. The main obstacle to the digital cancellation is a nonlinearity caused by the RF imperfection. The authors in~\cite{nonlinear1, nonlinear2,nonlinear3,access} discussed nonlinear digital cancellation methods that handle the power amplifier's (PA) nonlinearity, in-phase and quadrature (I/Q) imbalances, and phase noise.    
Generally, the nonlinear digital cancellation requires more computational overhead than its linear counterpart.

\begin{figure}[t]
	\begin{center}
		\resizebox{3in}{!}{\includegraphics{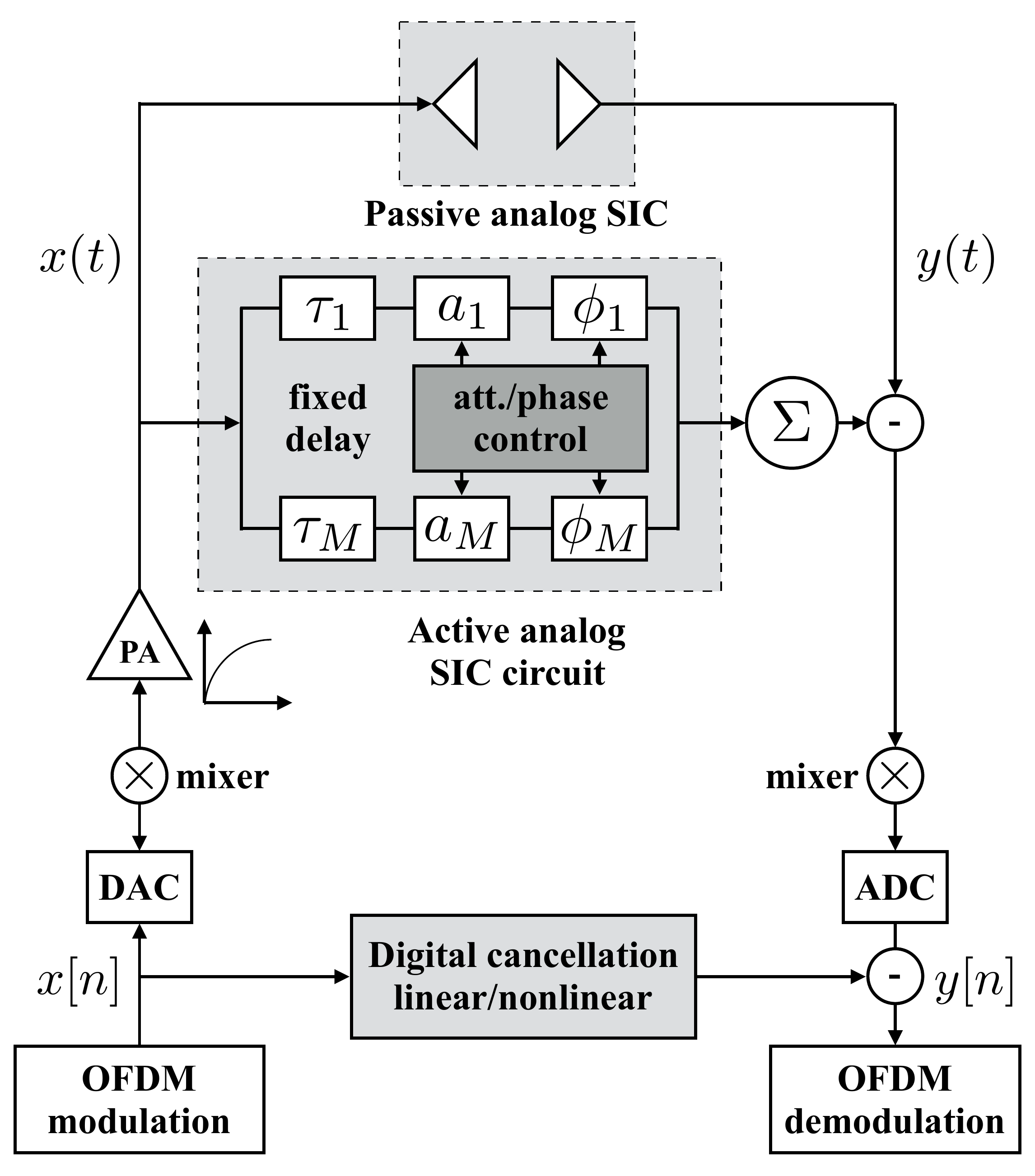}}
		\caption{A block diagram of the self-interference cancellation simulator.}	
		\label{bloack_diagram}
	\end{center}
\end{figure}

In this paper, we build an OFDM-based analog-digital self-interference cancellation simulator (Fig.~\ref{bloack_diagram}) that directly reflects the results of analog cancellation in the digital cancellation.
Considered in the simulation are such practical limitations such as PA's nonlinearity, the analog-to-digital converter (ADC) dynamic range. 
Through extensive simulations, we can design an efficient full-duplex system by analyzing the hardware and computational complexity and the performance of cancellation.

The rest of this paper is organized as follows. In Section~\ref{sec_2}, we discuss the hurdles of the SIC. In Section~\ref{sec_3}, we introduce our simulation methodology of the self-interference cancellation and also present adopted analog cancellation and digital cancellation methods. In Section~\ref{sec_4}, we present comparative performance evaluations of the adopted SIC methods in the OFDM system. Finally, in Section~\ref{sec_5}, we present our conclusions.

\section{Self-interference cancellation}
\label{sec_2}

\begin{figure}
	\begin{center}
		\resizebox{3in}{!}{\includegraphics{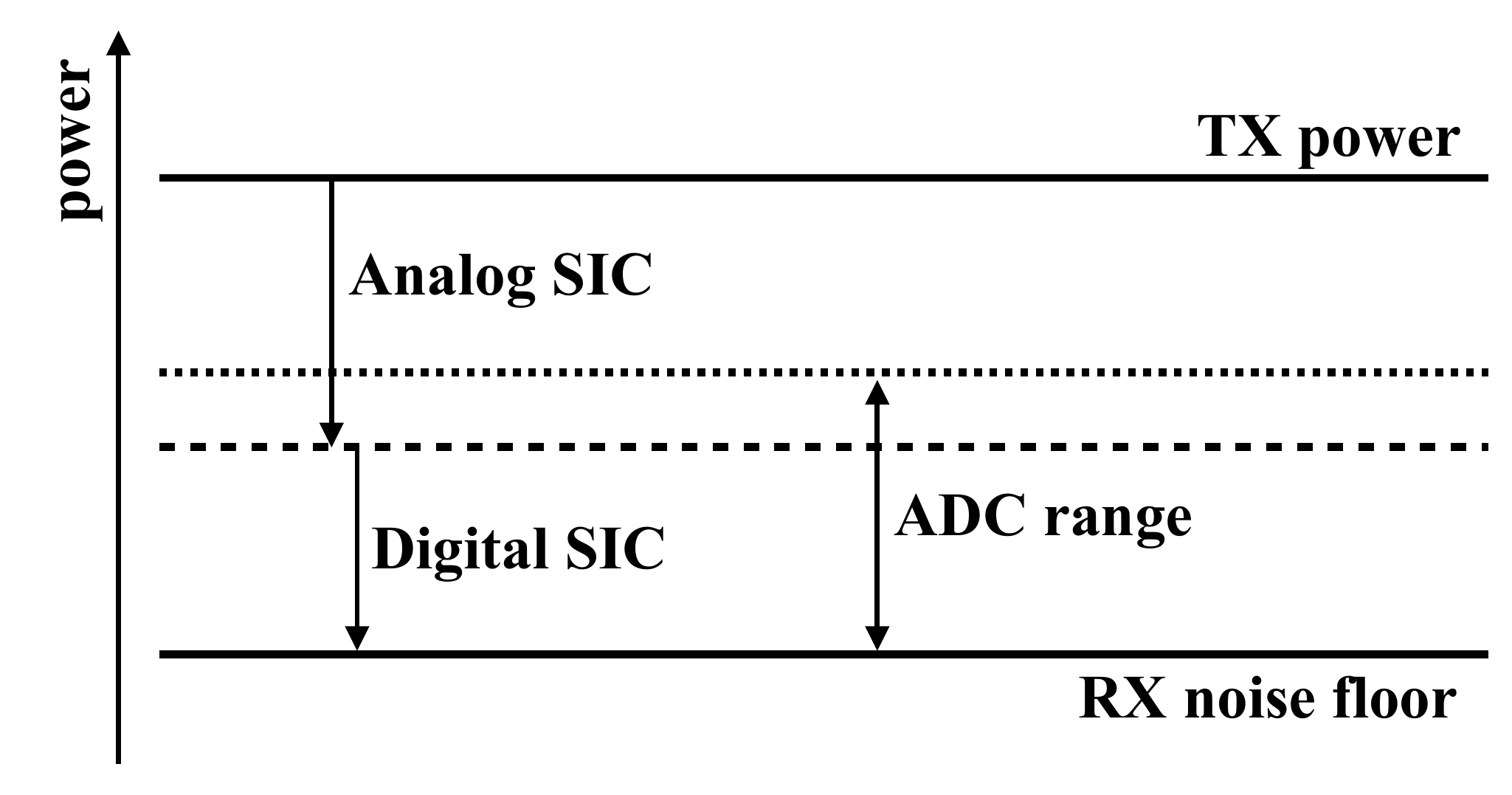}}
		\caption{The relation between power levels and SIC performances in a full-duplex system.}	
		\label{dynamic_range}
	\end{center}
\end{figure}

The most significant hurdle for SIC is the fact that an ADC has to convert the SI and signal-of-interest simultaneously. It is possible only if the SI is sufficiently removed in the analog domain to fall within the dynamic range of the ADC. 
Fig.~\ref{dynamic_range} illustrates an example of the relation between power levels and SIC performances in a full-duplex system. Moreover, transmitter noises must be suppressed to the receiver noise floor in the analog domain since it cannot be handled by digital cancellation.
In the typical full-duplex system, it is not possible to satisfy the ADC requirement with the passive analog cancellation alone~\cite{midu,dual-polar}. In a single antenna system, for instance, a circulator is widely used as an isolator. A circulator is a 3-port device that steers the signal entering any port, transmitting it to the next port only~\cite{Sachin2013Full}. When a signal enters to Port 1, the ferrite changes a magnetic resonance pattern to create a null at  Port 3. It is known that a typical isolation value of an RF circulator is approximately 15 to \SI{20}{\deci\bel}. 

Recently, to satisfy the requirements mentioned above, an adaptive circuit-based active analog cancellation is adopted in a full-duplex system. 
The authors in~\cite{Sachin2013Full} proposed an adaptive circuit composed of fixed delays and variable attenuators; these regenerate and subtract the leakages from the circulator. 
The leakages consist of the following three components: 1) direct leakage from Port 1 to Port 3, 2) reflection from the antenna at Port 2, and 3) reflection from objects near the transceiver. 
Theoretically, this requires an infinite number of fixed delay lines to completely mitigate the SI that has an actual unknown delay. Due to practical issues, the authors in~\cite{Sachin2013Full} used eight different fixed delay lines to cancel out each main leakage component (i.e., direct leakage and reflection from the antenna). In this case, a required size of the circuit must be 10$\times$\SI{10}{\centi\meter}, which is not implementable in small mobile devices.
Not only does the size of the circuit linearly increase as the number of taps increased, but the required computational overhead for tuning also increases. 
A structure and tuning algorithm of the adaptive circuit should be designed considering both the SIC performance and the system requirements.



\section{Simulation of the self-interference cancellation}

In this section, we introduce how each block of the simulator, shown in Fig.~\ref{bloack_diagram} is modeled. First, we represent the SI channel using a tapped-delay-line (TDL) model. Passive analog cancellation is modeled as the taps that have the attenuation and the delay of the leakage. A fixed delay line in the active analog cancellation is modeled in the same way, where the amplitude and the phase of the tap are controlled by a tuning algorithm in the baseband. We then transform the SI channel into a baseband equivalent form. We adopt a baseband modeling of the nonlinear PA, DAC, and ADC. 

\label{sec_3}
\begin{table}
	\caption{Simulation Parameters}
	\label{table_3}
	\begin{center}
		\begin{tabular}{ccc}
			\hline\hline
			System Parameter &Notation &Values \\
			\hline
			Center Frequency&$f_\text{c}$ &\SI{1.5}{\giga\hertz}\\
			Bandwidth &$B$&\SI{20}{\mega\hertz}\\
			FFT Size &&64\\
			Used Subcarrier &&52\\
			CP Length &&16\\
			ADC Resolution &&10, 14, $\infty$\\
			TX Power & &5 dBm \\
			PA Gain&&15dB\\
			$\rm{P}_\text{1dB}$&&5.64 dBm, 23.17 dBm\\ 
			Receiver Noise Floor & &-80 dBm \\
			Power Initialization && $10^{-5}$\\
			\hline \hline
		\end{tabular}
	\end{center}
\end{table}
\subsection{Passband simulation}
In the simulation of analog SIC, we have to generate the leakage delays from passive analog SIC. 
A time-invariant passband channel $h_{\text{p}}(t)$ is represented as 
\begin{equation}
\label{eq.h_p}
h_{\text{p}}(t)=\sum_{i=0}^{L}c_{\text{p}}^i\delta(t-\tau_i),
\end{equation} 
where $L+1$ is the number of taps, $c_{\text{p}}^i$ is a gain of the $i$-th tap, and $\tau_i$ is a delay of the $i$-th tap.
We model a circulator with two delayed taps for the rest of this paper.
In the simulation, the direct leakage has \SI{15}{\deci\bel} attenuation and \SI{300}{\pico\second} delay, and the reflection from the antenna has \SI{17.5}{\deci\bel} attenuation and \SI{3}{\nano\second} delay. A problem occurs when the delay is not a multiple of the simulation period.
We can resolve this problem by choosing a simulation period as a divisor of all the delays we want to generate in the simulation. In practice, however, the delay of each leakage is a continuous random variable in practice, which is impossible to be represented as a multiple of the simulation period. 

\begin{figure}
	\begin{center}
		\resizebox{3in}{!}{\includegraphics{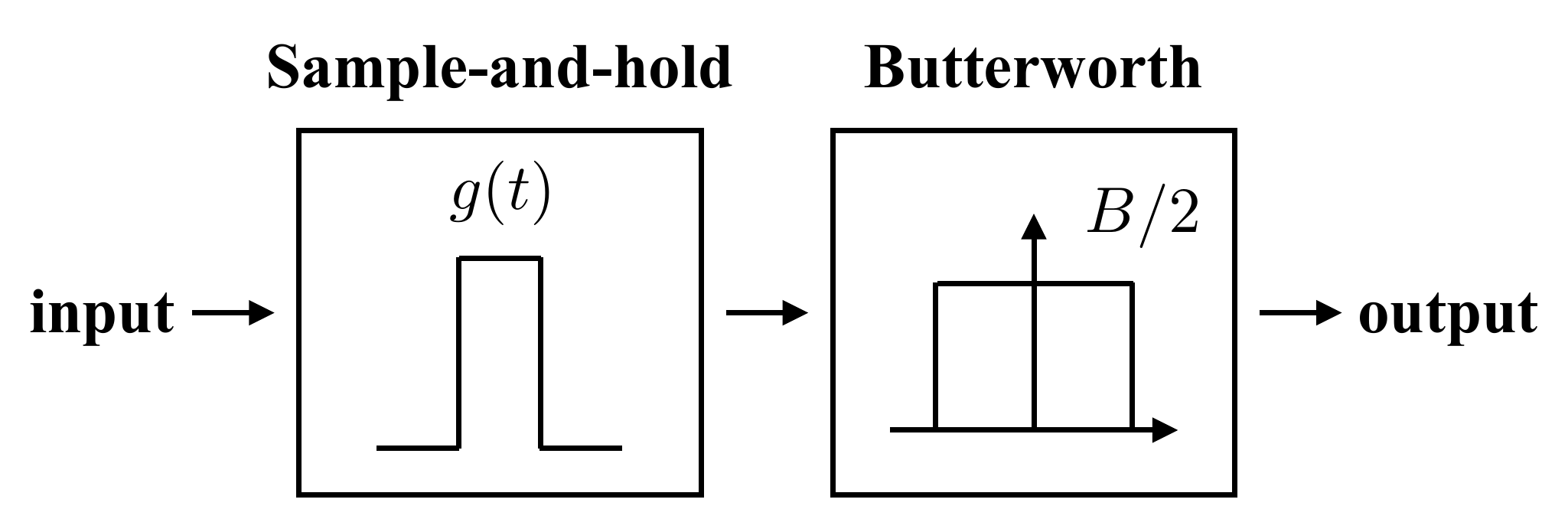}}
		\caption{A block diagram of a DAC emulation applied in the simulator.}	
		\label{DAC_diagram}
	\end{center}
\end{figure}

Typically, a passband channel is transformed into a baseband equivalent form to avoid the misaligned problem in the simulation~\cite{simulation}. The equivalent transformation is well-known and easy to calculate with an assumption of an ideal sinc interpolation in the digital-to-analog converter (DAC). Fig.~\ref{DAC_diagram} depicts a block diagram of the DAC in the simulation. 
An impulse response of sample-and-hold $g(t)$ is defined as
\begin{equation}
\label{pulse_shaping}
g(t)=
\begin{cases}
1, & \text{if $0\leq n\leq T$},\\
0, & \text{otherwise},
\end{cases}
\end{equation}
where $T$ is the baseband elementary period.
The reconstruction filter is the causal Butterworth filter of order 3 and a cut-off frequency of $B/2$, where $B$ is the bandwidth of the baseband signal. 

With this DAC setup, we calculate a baseband equivalent form of~\eqref{eq.h_p}. 
Let the impulse response of the given Butterworth filter be $u(t)$. A baseband equivalent form of the arbitrary passband channel tap is represented as 
\begin{align}
\label{eq.hb_equiv1}
{h}_{\text{b}}^{c_\text{p}^i,\tau_i}[n]=
\begin{cases}
0, & \text{if $0\leq n\leq \floor{\frac{\tau_i}{T}}$},\\
c_\text{p}^i\!\int_{0}^{T-\tau_i}u(t)dt, & \text{if $n=\floor{\frac{\tau_i}{T}}+1$}, \\
c_\text{p}^i\!\int_{(n-1-\floor{\frac{\tau_i}{T}})T-\tau_i}^{(n-\floor{\frac{\tau_i}{T}})T-\tau_i}u(t)dt, & \text{otherwise}, 
\end{cases} 
\end{align}
where $\pmb{h}_{\text{b}}^{c_\text{p}^i,\tau_i}=[h_{\text{b}}^{h_\text{p}^i,\tau_i}[0],h_{\text{b}}^{h_\text{p}^i,\tau_i}[1],\cdots]$ is the baseband equivalent form of the passband channel tap $c_{\text{p}}^i\delta(t-\tau_i)$. 
Summing all the taps in the passband channel~\eqref{eq.h_p}, we get the baseband equivalent channel $\pmb{{h}}_{\text{b}}$ as 
\begin{equation}
\label{eq.h_b}
\pmb{{h}}_{\text{b}}=\sum_{i=0}^{L}\pmb{h}_{\text{b}}^{c_\text{p}^i,\tau_i}. 
\end{equation} 
Table~\ref{table_tap} depicts the simulated passband tap delays and attenuations.


\begin{table}
	\caption{Simulated passband tap delays and attenuations}
	\label{table_tap}
	\begin{center}
		\begin{tabular}{ccc}
			\hline\hline
			Source &Attenuation &Delay \\
			\hline
			Direct Leakage &\SI{-15}{\deci\bel}&\SI{300}{\pico\second}\\
			\hline
			Reflection from antenna&\SI{-17.5}{\deci\bel}&\SI{3}{\nano\second}\\
			\hline
			Reflection from objects &\SI{-60}{\deci\bel}&\SI{20}{\nano\second}\\
			Reflection from objects &\SI{-90}{\deci\bel}&\SI{60}{\nano\second}\\
			Reflection from objects &\SI{-100}{\deci\bel}&\SI{90}{\nano\second}\\
			Reflection from objects &\SI{-100}{\deci\bel}&\SI{120}{\nano\second}\\
			
			\hline 
		\end{tabular}
	\end{center}
\end{table}

\subsection{Active analog cancellation simulation}
For the simulation of active analog cancellation we adopt a simple adaptive circuit. It consists of fixed delay lines with variable attenuators and phase shifters. Values of the attenuators and phase shifters are tuned via solving
\begin{equation}
\label{eq.h_circuit}
\underset{a_1,..,a_M,\phi_1,..,\phi_M}{\text{min}}   \left(\pmb{H}-\sum_{i=1}^{M}a_ie^{j\phi_i}\pmb{{H}}^{i}\right)^2,
\end{equation} 
where $a_i,\phi_i$ is the attenuation and phase shifter value of the $i$-th delay line, $\pmb{H}$ is the frequency domain representation of $\pmb{h}_{b}$, and $\pmb{{H}}^{i}$ is the frequency response of the $i$-th delay line without attenuation and phase shifting. The authors in~\cite{lincoln} theoretically modeled $\pmb{{H}}^{i}$ and then formulated~\eqref{eq.h_circuit} as a convex problem. Instead, we obtain $\pmb{{H}}^{i}$ through the following steps: 1) Set $h_\text{p}(t)$ as
\begin{equation}
\label{eq.h_cir_measure}
h_\text{p}(t)=\delta(t-d_i),
\end{equation} 
where $d_i$ is the fixed delay value of $i$-th delay line.
2) Calculate the baseband equivalent channel using~\eqref{eq.h_b}.
3) Send a pilot and estimate the channel in the frequency domain
to obtain $\pmb{{H}}^{i}$.
Using a precalculated frequency response of the circuit (i.e., $\{\pmb{{H}}^{i}\}$), we can solve~\eqref{eq.h_circuit} by the least square method.
During the initial stage, the pilot is transmitted with little power to avoid the ADC saturation without the active analog cancellation.

\begin{figure}[t]
	\begin{center}
		\includegraphics[width=0.98\columnwidth,keepaspectratio]{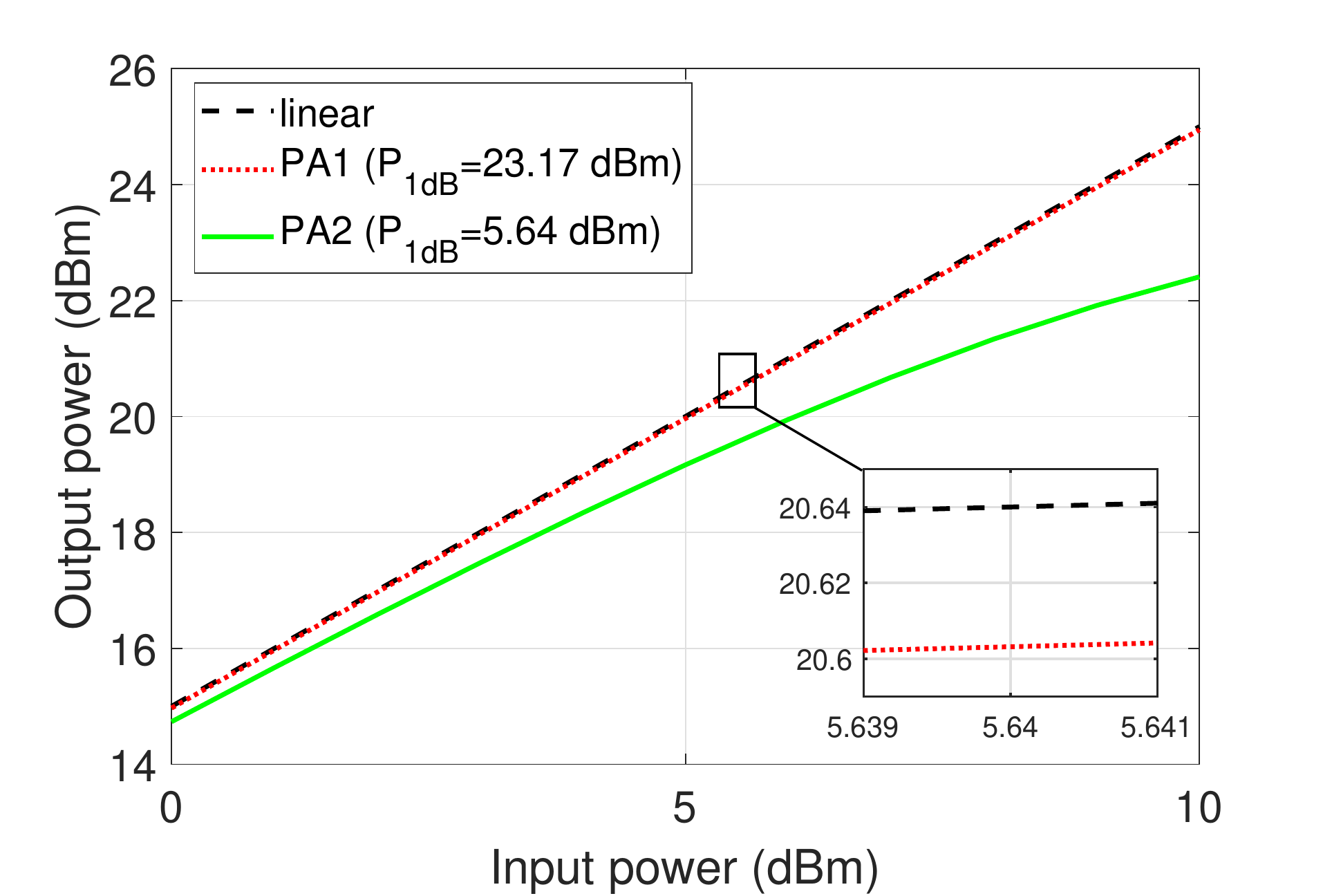}
		\caption{The characteristics of power amplifiers used in the simulation.}	
		\label{p1dB}
	\end{center}
\end{figure}
\subsection{Digital cancellation simulation}

To simulate digital cancellation, we consider both the linear and nonlinear digital cancellation. Linear digital cancellation assumes that the residual SI channel is linear.
A frequency response of the channel at the $k$-th subcarrier can be estimated as 
\begin{equation}
\hat{H}[k]=\frac{Y[k]}{X_\text{p}[k]},
\end{equation}
where $X_{\text{p}}[k]$ and $Y[k]$ are the pilot and received signals at the $k$-th subcarrier. 
The residual SI is then reconstructed as 
\begin{equation}
\hat{S}[k]=\hat{H}[k]X[k],
\end{equation}
where $X[k]$ is the known transmitted signal at the $k$-th subcarrier. Then the reconstructed SI is subtracted from the received signal.

\begin{table*}[t]
	\caption{Comparative performance evaluation of the adopted analog and digital cancellation methods. }
	\label{table_results}
	\begin{center}
		\begin{tabular}{c|c|c|c|c}
			\hline\hline 
			Power Levels&Adaptive Circuit& 10-bit ADC & 14-bit ADC&$\infty$-bit ADC \\
			\hline
			
			&&Analog: \SI{49.67}{\deci\bel} &Analog: \SI{49.68}{\deci\bel}&Analog: \SI{49.77}{\deci\bel}\\
			TX Power: 5 dBm&Single-tap &Linear Digital: \SI{44.07}{\deci\bel}&Linear Digital: \SI{44.50}{\deci\bel}&Linear Digital: \SI{46.52}{\deci\bel}\\
			PA1  &(\SI{1.5}{\nano\second})&Nonlinear Digital: \SI{47.04}{\deci\bel}&Nonlinear Digital: \SI{46.80}{\deci\bel} &Nonlinear Digital: \SI{47.76}{\deci\bel}  \\
			\cline{2-5}
			($\rm{P}_\text{1dB}$: 23.17 dBm, Gain: 15 dB)&&Analog: \SI{60.85}{\deci\bel}&Analog: \SI{70.60}{\deci\bel}&Analog: \SI{72.44}{\deci\bel}\\
			Receiver Noise: -80 dBm&Double-tap&Linear Digital: \SI{33.33}{\deci\bel}
			&Linear Digital: \SI{23.70}{\deci\bel}&Linear Digital: \SI{24.06}{\deci\bel}\\
			&(\SI{0.5}{\nano\second}, \SI{2.5}{\nano\second})&Nonlinear Digital: \SI{35.70}{\deci\bel}&Nonlinear Digital: \SI{26.12}{\deci\bel}  &Nonlinear Digital: \SI{25.04}{\deci\bel} \\
			\hline\hline
			
			&&Analog: \SI{45.39}{\deci\bel} &Analog: \SI{45.46}{\deci\bel}&Analog: \SI{44.69}{\deci\bel}\\
			TX Power: 5 dBm&Single-tap &Linear Digital: \SI{16.85}{\deci\bel}&Linear Digital: \SI{20.11}{\deci\bel}&Linear Digital: \SI{20.17}{\deci\bel}\\
			PA2 &(\SI{1.5}{\nano\second})&Nonlinear Digital: \SI{35.08}{\deci\bel}&Nonlinear Digital: \SI{50.39}{\deci\bel} &Nonlinear Digital: \SI{51.73}{\deci\bel}  \\
			\cline{2-5}
			($\rm{P}_\text{1dB}$: 5.64 dBm, Gain: 15 dB)&&Analog: \SI{45.59}{\deci\bel}&Analog: \SI{45.36}{\deci\bel}&Analog: \SI{45.37}{\deci\bel}\\
			Receiver Noise: -80 dBm &Double-tap&Linear Digital: \SI{16.97}{\deci\bel}
			&Linear Digital: \SI{20.35}{\deci\bel}&Linear Digital: \SI{20.34}{\deci\bel}\\
			&(\SI{0.5}{\nano\second}, \SI{2.5}{\nano\second})&Nonlinear Digital: \SI{34.85}{\deci\bel}&Nonlinear Digital: \SI{50.46}{\deci\bel} &Nonlinear Digital: \SI{51.57}{\deci\bel}\\
			\hline\hline	
		\end{tabular}
	\end{center}
\end{table*}

The limitation of linear digital cancellation is nonlinear SI components, which mainly come from the PA's nonlinearity. The parallel Hammerstein model~\cite{hammerstein,digital} is widely used to describe the nonlinearity as
\begin{equation}
\label{eq.pa}
x_{\text{PA}}[n]=\sum_{k=0}^{K-1}\sum_{p=0}^{P-1}\psi_{k,p}\left|x[n-p]\right|^{2k}x[n-p],
\end{equation}
where $x[n]$ and $x_{\text{PA}}[n]$ are the time domain input and output signals on time $n$, $2K\!-\!1$ is the highest order of the model, $P$ is the number of the model's taps, and $\{\psi_{k, p}\}$ are the nonlinear coefficients.
The received signal is then modeled as 
\begin{equation}
\label{eq.pa}
y[n]=\sum_{k=0}^{K-1}\sum_{\ell=0}^{M+P-1}b_{k,\ell}|x[n-\ell]|^{2k}x[n-\ell]+z[n],
\end{equation} 
where $y[n]$ and $z[n]$ are the received signal and noise at time~$n$, $M$ is the number of baseband equivalent channel taps, and $\{b_{k,\ell}\}$ are the effective nonlinear coefficients of the channel.
Fig.~\ref{p1dB} compares the characteristics of the PA implemented in the simulation compare to the linear PA.
Each PA is modeled as 
\begin{align}
\label{eq.pa_simul}
\begin{split}
x_{\text{PA1}}[n]&=x[n]-10^{-4}|x[n|]^3-10^{-7}|x[n]|^5, \\
x_{\text{PA2}}[n]&=x[n]-10^{-2}|x[n|]^3-10^{-7}|x[n]|^5.
\end{split}
\end{align}
Similar to the linear digital cancellation, $\{b_{k,\ell}\}$ are estimated using  pilot signals~\cite{asilomar}. The residual SI is then reconstructed and subtracted.

\section{Performance evaluations and discussions}
\label{sec_4}
In this section, we provide the following numerical analysis:
\begin{itemize}
\item Compare the performance of the adaptive circuit with a different configuration (i.e., the number of delay lines and fixed delay values).
\item Performance evaluation of the adopted analog and digital cancellation method with different ADC resolution and power amplifier nonlinearity. 
\end{itemize}

\begin{figure}[t]
	\begin{center}
		\includegraphics[width=0.98\columnwidth,keepaspectratio]
		{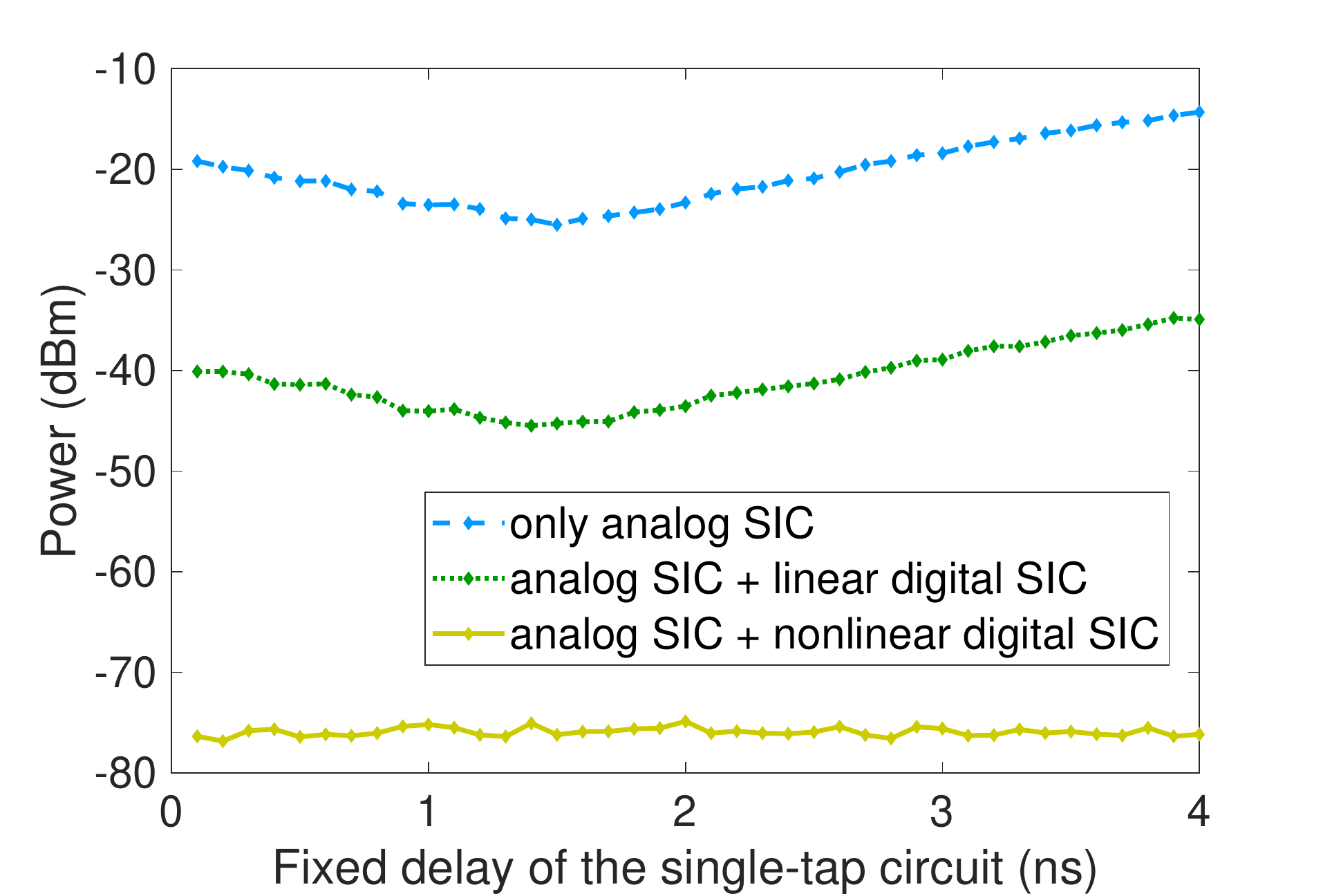}
		\caption{Comparison of the SIC performance by changing the fixed delay of the single-tap adaptive circuit. $\rm{P}_\text{1dB}$ and ADC resolution is set to 5.64 dBm and 14-bit. }	
		\label{power_level}
	\end{center}
\end{figure}

Fig.~\ref{power_level} shows a comparison of the single-tap adaptive circuit with different fixed delay value. Analog cancellation amount of the single-tap circuit is maximized when the fixed delay is \SI{1.5}{\nano\second}. A red line depicts the results of the analog and linear digital cancellation. In this case, the overall SIC performance is dependent on the analog cancellation, as the nonlinear SI components are mitigated by the analog cancellation only.
A yellow line depicts the results of the analog and nonlinear digital cancellation. The overall SIC performance reveals, in this case, a certain tendency. This is because of the parallel Hammerstein model fully describes the system adopted in the simulation, aside from the quantization noise in ADC. In practice, transmitter noise will still remain if it is not suppressed to the receiver noise floor in the analog domain. 

Table~\ref{table_results} shows the cancellation amounts of the adopted analog and digital cancellation methods. 
We compare the double-tap (50 ns, 250 ns) adaptive circuit to the single-tap (150~ns) adaptive circuit. Intuitively, the double-tap adaptive circuit shows better analog cancellation performance in the case of PA1, whose $\rm{P}_\text{1dB}$ is 23.17 dBm. However, the performances of the single-tap and the double-tap canceller are saturated in the case of PA2, whose $\rm{P}_\text{1dB}$ is 5.64 dBm. 
It is because the tuning algorithm is based on the linear SI channel estimation. 
Although the adaptive circuit can mitigate the nonlinear SI component since the circuit tapped the transmitted signal as an input, it still assumes that the SI channel is linear.
Therefore, the performance of the active analog cancellation deteriorates as the PA nonlinearity increases.  

The linear digital cancellation shows better performance in the case of PA1, which has lower nonlinearity than the PA2. 
As noted in Section~\ref{sec_2}, the ADC dynamic range set a minimum goal for the analog cancellation. 
For instance, at least a 40 dB of analog cancellation must be achieved in the case of a 10-bit ADC since the dynamic range of 10-bit ADC is about 60 dB. 
The simulation results show that it is insufficient due to the high peak-to-average power ratios (PAPR) of OFDM signals. Even though the analog cancellation amount is approximately 45 dB in the case of 10-bit ADC with PA2, digital cancellation has poor performances because some of the OFDM signals are clipped. A 14-bit ADC provides sufficient dynamic range for all the cases in the simulation. Therefore, the adopted cancellation methods have the almost same performances with the $\infty$-bit ADC cases.

\section{Conclusion}
\label{sec_5}
In this paper, we have described a simulator that evaluates the analog-digital integrated SIC performances in the OFDM system. For a realistic simulation, we reflect the practical characteristics of PA, DAC, and ADC.  
Performance of the adopted circuit is analyzed in the various system environments and circuit configurations.
Through the simulation, we can configure an efficient full-duplex system. For  future work, we will consider an extension to a MIMO full-duplex configuration~\cite{MIMO, MIMO2, MIMO3}.

\section{Appendix}
Let $x[n]$ be the DAC input baseband signals at the $n$-th time slot. The output signals of the sample-and-hold block in the DAC, $x_{\text{sh}}(t)$ is represented as 
\begin{equation}
\label{dac_sh}
x_\text{sh}(t)=g(t)*\sum_{n=0}^{N}x[n]\delta(t-nT),
\end{equation}
where $N$ is the length of the transmitted baseband signal.
After the low-pass filtering, the ouput signals of the DAC, $x_\text{out}(t)$ is then represented as  
\begin{align}
\label{dac_output}
\begin{split}
x_\text{out}(t)&=x_\text{sh}(t)*u(t)\\
			   &=\sum_{n=0}^{N}x[n]\delta(t-nT)*(g(t)*u(t)) \\
			   &=\sum_{n=0}^{N}x[n]\delta(t-nT)*\left(\int_{t-T}^{t}u(\tau)d\tau\right)  \\
			   &=\sum_{n=0}^{N}x[n]\int_{t-nT-T}^{t-nT}u(\tau)d\tau.
\end{split}
\end{align}
Then the passband signal $x_{\text{pass}}$ is
\begin{align}
\label{mixer_output}
\begin{split}
x_\text{pass}(t)&=e^{j2\pi f_ct}x_\text{out}(t)\\
&=e^{j2\pi f_ct}\sum_{n=0}^{N}x[n]\int_{t-nT-T}^{t-nT}u(\tau)d\tau.
\end{split}
\end{align}
Finally, the received baseband signal is obtained as
\begin{align}
\label{y_n}
\begin{split}
y[m]&=\int_{\infty}^{\infty}e^{-j2\pi f_c\tau}x_{\text{pass}}(\tau)h_{\text{p}}(mT-\tau)d\tau\\
&=\int_{\infty}^{\infty}e^{-j2\pi f_c\tau}x_{\text{pass}}(\tau)c_{\text{p}}^i\delta(mT-\tau-\tau_i)d\tau\\
&=c_{\text{p}}^ie^{-j2\pi f_c(mT-\tau_i)}x_{\text{pass}}(mT-\tau_i)
\end{split}\\
&=c_{\text{p}}^i\sum_{n=0}^{N}x[n]\int_{mT-\tau_i-(n+1)T}^{mT-\tau_i-nT}u(\tau)d\tau,
\end{align}
where $y[m]$ is the received baseband signal at $m$-th time slot.
Now the received baseband signal is represented as a linear combination of the transmitted baseband signal. In the case of causal Butterworth filter, we get~\eqref{eq.hb_equiv1}.



\vspace{12pt}

\end{document}